\documentclass[prd,twocolumn]{revtex4}
\usepackage{dcolumn}
\usepackage{multirow}
\usepackage{graphicx}
\usepackage{amssymb}
\usepackage{bm}
\usepackage{hyperref}
\usepackage{epstopdf}
\usepackage{color}
\usepackage{mathrsfs}
\usepackage{amsmath,amssymb,amsthm}
\usepackage{rotating}
\usepackage{sverb, longtable}
\usepackage{subfigure}

\usepackage[graphicx]{realboxes}
\usepackage{adjustbox}
\begin{document}

\title{Testing the black hole area law with Event Horizon Telescope}
\author{Deng Wang}
\email{cstar@nao.cas.cn}
\affiliation{National Astronomical Observatories, Chinese Academy of Sciences, Beijing, 100012, China}
\begin{abstract}
Hawking's black hole area theorem can be tested by monitoring the evolution of a single black hole over time. Using current imaging observations of two supermassive black holes M87* and Sgr A* from the Event Horizon Telescope (EHT), we find their horizon area variation fractions are consistent with the prediction of the black hole area law at the $1\,\sigma$ confidence level. We point out that whether the black hole area law is valid or not could be determined by future high precision EHT observations of Sgr A*.

\end{abstract}
\maketitle

\section{Introduction}
General relativity (GR) has been tested by using a large number of observations ranging from large scales to very small scales \cite{Will:2014kxa}, since Einstein proposed it in 1915 \cite{Einstein1915}. For example, cosmological observations leading to the discovery of the late-time cosmic acceleration give strong constraints on the possible deviations from GR at both the background and perturbation levels \cite{Ferreira:2019xrr}. In our solar system, astrophysical observations such as the precession of Mercury's orbit \cite{Verma:2013ata}, the deflection of a light ray when solar eclipses occur \cite{Bertotti:2003rm} and the gravitational time delays when photons pass the solar surface \cite{Lambert:2011}, have verified the success of GR to a high precision.

A black hole as the most important compact object predicted by GR, plays a basic role in fundamental physics. Since investigated by Schwarzschild for the first time \cite{Schwarzschild1916}, black holes always lie in the core of theoretical physics when one tries to unify GR with quantum mechanics. The characteristic of a black hole is that it has an event horizon where a light ray can not escape from the tight binding of strong gravity. One of the most intriguing aspects of black holes is the second law of black hole mechanics, i.e., the so-called Hawking's area theorem \cite{Hawking:1971tu}, which states that the total horizon area of a black hole never decreases over time. This is a basic result from GR and the cosmic censorship conjecture \cite{Penrose:1969pc}. It is natural that one can check the validity of GR by testing the black hole area law. In general, there are two approaches to achieve this goal. The former is measuring the area variation between two progenitor black holes and the remnant black hole for a gravitational wave burst from a binary black hole merger \cite{Isi:2020tac}, while the latter is monitoring a specific black hole for a long time and measuring its area variation. For the former case, current gravitational wave observations from the LIGO and VIRGO collaboration \cite{LIGOScientific:2016aoc,LIGOScientific:2016lio,LIGOScientific:2019fpa,LIGOScientific:2020tif} can help carry out the test. In \cite{Isi:2020tac}, the authors have confirmed the Hawking's area theorem by calculating the area variation between inspiralling and ringdown phases for the first LIGO's detection, GW150914. However, the second approach is still  unexplored until today. Fortunately, in 2019, the EHT collaboration firstly released four shadow images of the supermassive black hole M$87^*$ residing in the giant elliptical galaxy Messier 87 \cite{EventHorizonTelescope:2019dse,EventHorizonTelescope:2019ggy,EventHorizonTelescope:2019pgp,EventHorizonTelescope:2019ths} based on observations from April 5 to April 11 in 2017. Subsequently, in 2022, they released two shadow images of the supermassive black hole Sgr A* \cite{EventHorizonTelescope:2022wok,EventHorizonTelescope:2022xnr,EventHorizonTelescope:2022xqj} in the Galactic center based on data from April 6 to April 7 in 2017. These imaging observations can provide two independent black hole systems to help implement the test. In this study, we attempt to test the black hole area law in light of the EHT imaging data. We find that Hawking's black hole area theorem is compatible with the EHT imaging observations at the $1\,\sigma$ confidence level.        

This study is organized as follows. In the next section, we describe the analysis methodology. In Section III, we display the data used. In Section IV, we exhibit the numerical results. The discussions and conclusions are presented in the final section.  

\begin{figure*}
	\centering
	\includegraphics[scale=0.35]{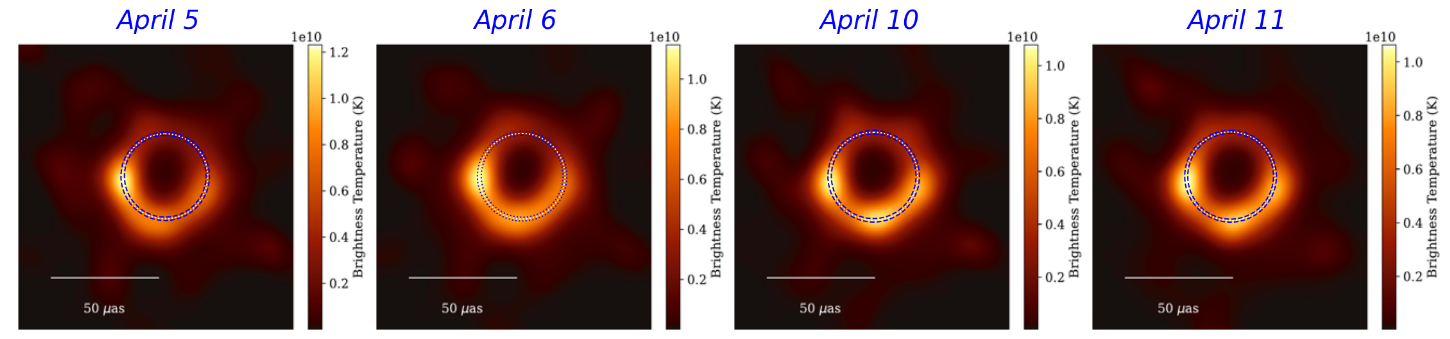}
	\caption{The EHT images of M87* from the eht-imaging parameter survey for April 5, April 6, April 10 and April 11 in 2017. The dotted (white) and blue circles are the mean values and $1\,\sigma$ limits of angular ring diameters, respectively. In principle, one can test the black hole area law through measuring the angular ring diameters at different times with a high precision.}\label{f1}
\end{figure*}

\begin{table*}[!t]
	\renewcommand\arraystretch{1.5}
	\caption{The $1\,\sigma$ confidence ranges of the relative area variation $\Delta A/A_0$ in different time intervals for M87* and Sgr A*. Here we also display the forecasted constraint on $\Delta A/A_0$ for Sgr A* by increasing the measurement precision of the angular ring diameter by 1 order of magnitude.}
	\setlength{\tabcolsep}{5mm}{
		{\begin{tabular}{@{}c|c|c|c|c|c@{}} \toprule
				Black holes                 & \multicolumn{3}{c}{M87*}      &   \multicolumn{2}{|c}{Sgr A*}                              \\
				\hline
				Time intervals            &April 5 to 6           &April 6 to 10      &April 10 to 11   &April 6 to 7           &Forecast                                    \\ \colrule
				$\Delta A/A_0$      &0.015$\pm$0.224  &0.060$\pm$0.230  &0.015$\pm$0.215  &0.118$\pm$0.232  &0.118$\pm$0.029           \\  
				\hline
				\hline
			\end{tabular}
			\label{t1}}}
\end{table*}

\begin{figure*}
	\centering
	\includegraphics[scale=0.55]{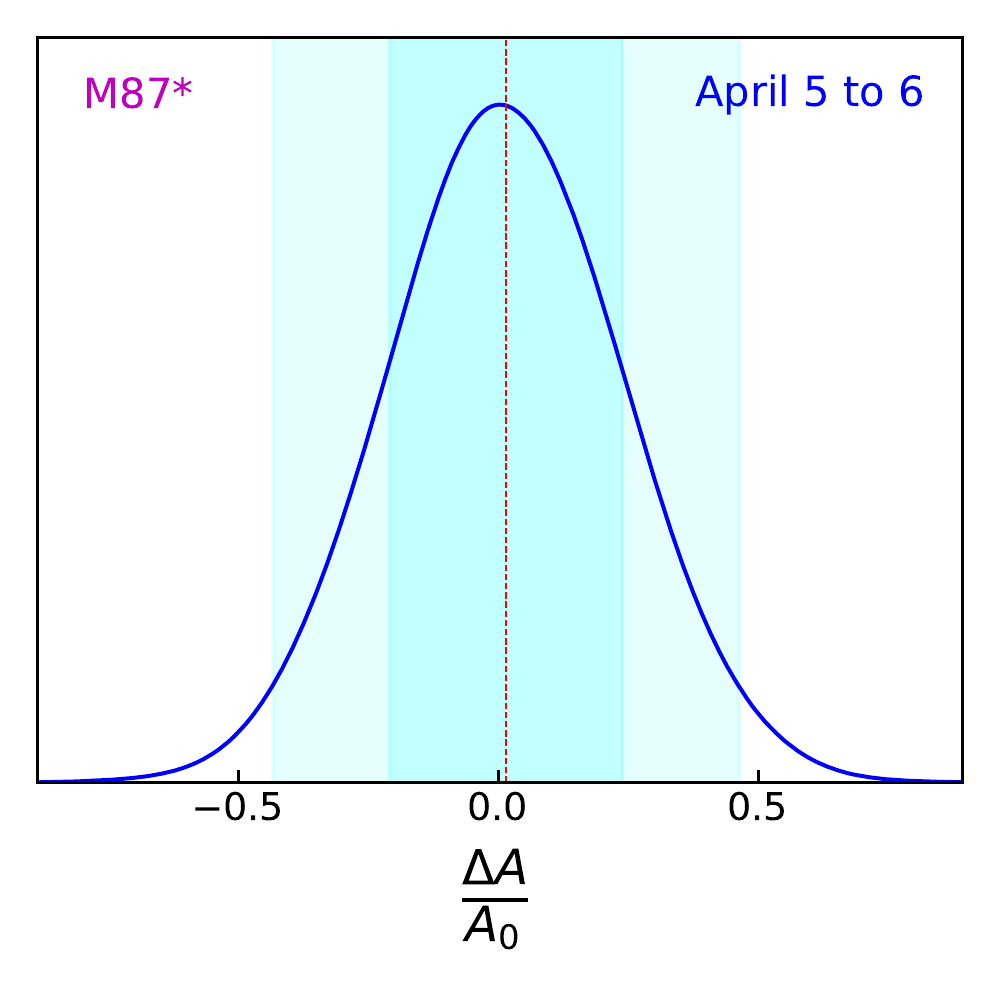}
	\includegraphics[scale=0.55]{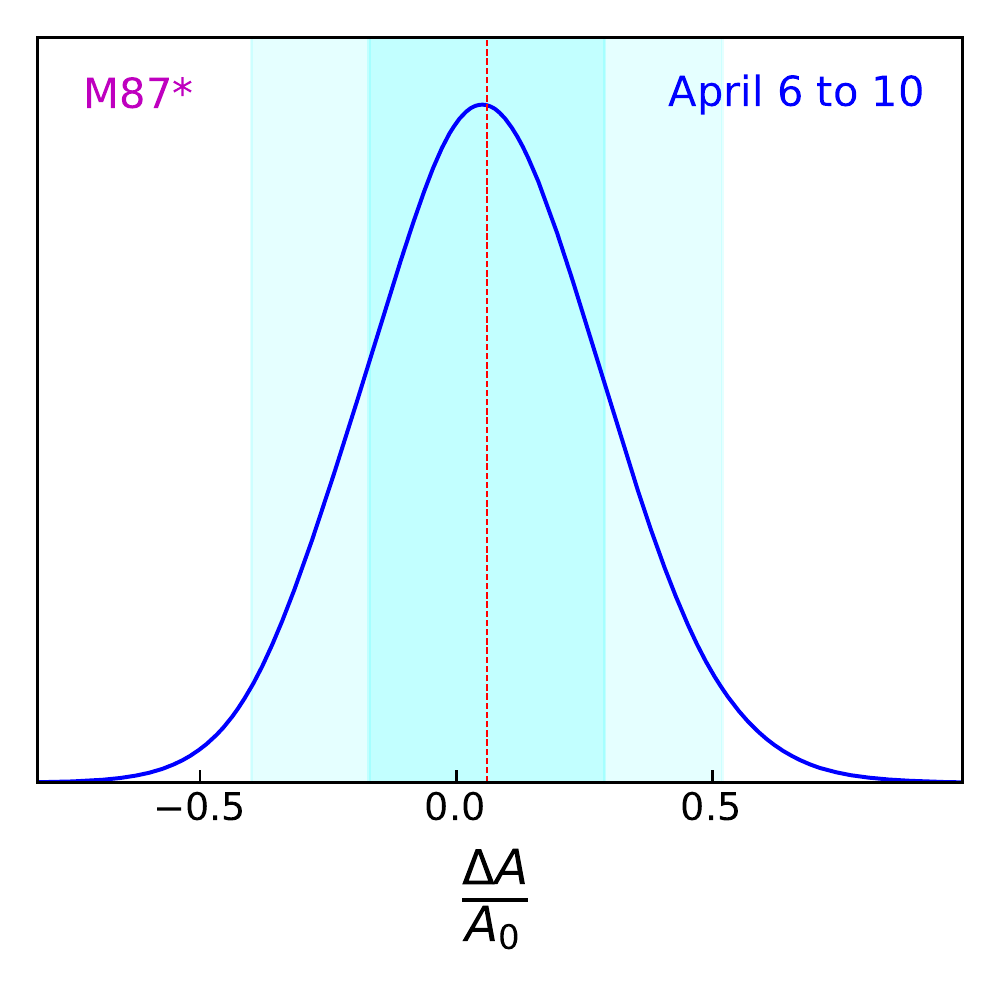}
	\includegraphics[scale=0.55]{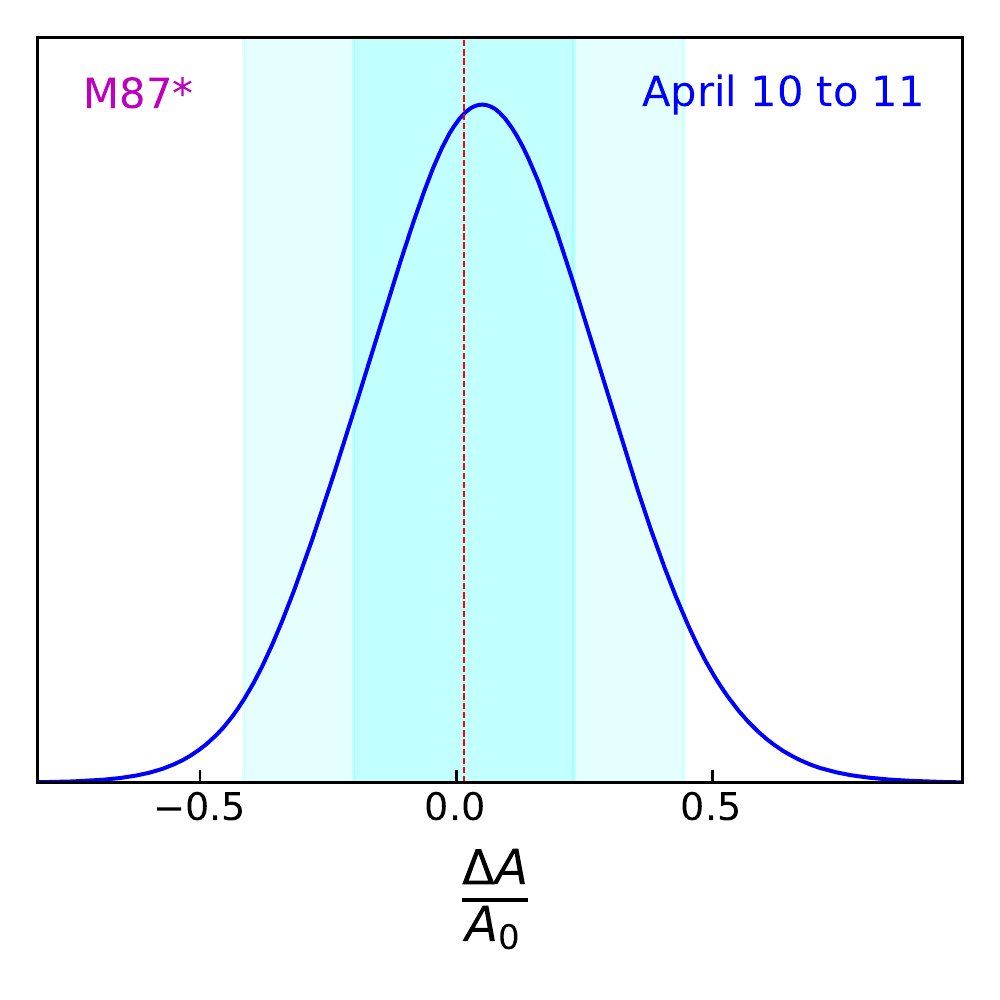}
	\caption{The 1-dimensional normalized posterior distributions of the area variation fraction $\Delta A/A_0$ of M87*. The vertical red line and cyan bands are the mean and $1\,\sigma$ and $2\,\sigma$ confidence ranges of $\Delta A/A_0$, respectively.}\label{f2}
\end{figure*}

\begin{figure}
	\centering
	\includegraphics[scale=0.55]{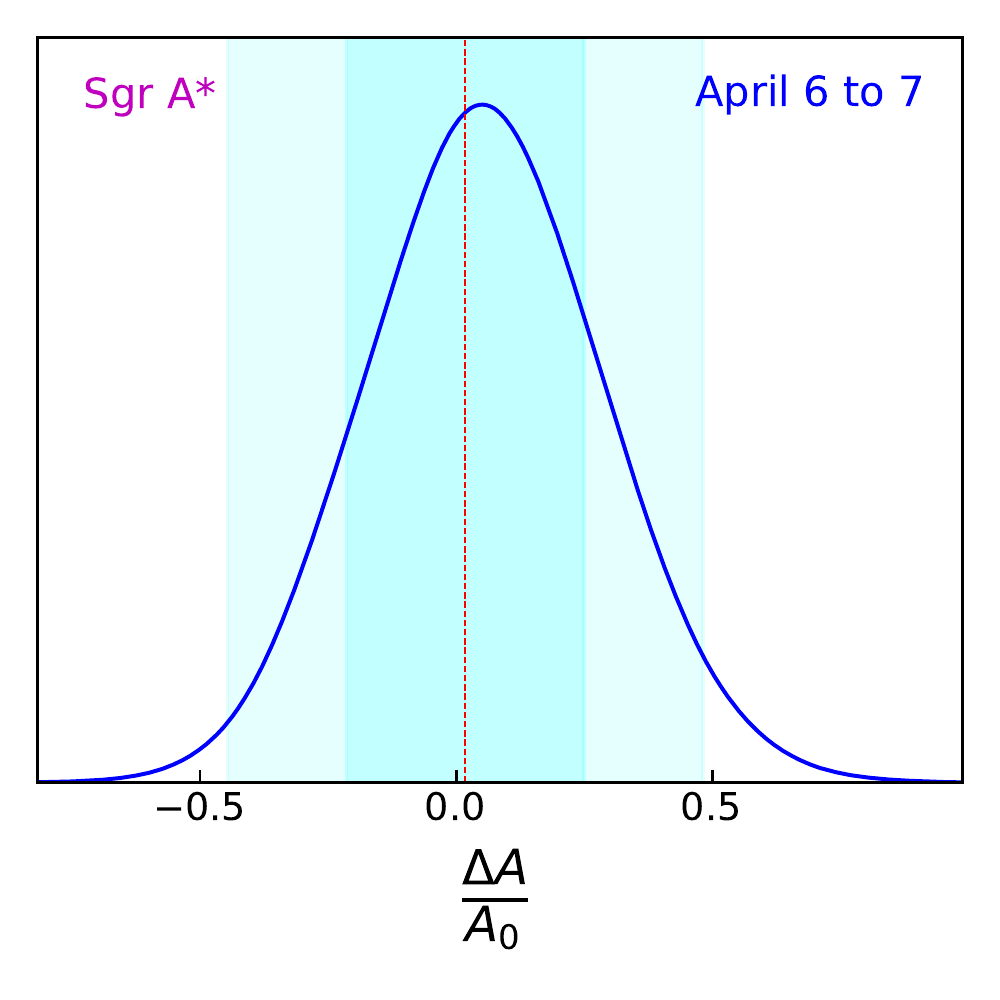}
	\caption{The 1-dimensional normalized posterior distributions of the area variation fraction $\Delta A/A_0$ of Sgr A*. The vertical red line and cyan bands are the mean and $1\,\sigma$ and $2\,\sigma$ confidence ranges of $\Delta A/A_0$, respectively.}\label{f3}
\end{figure}

\begin{figure}
	\centering
	\includegraphics[scale=0.55]{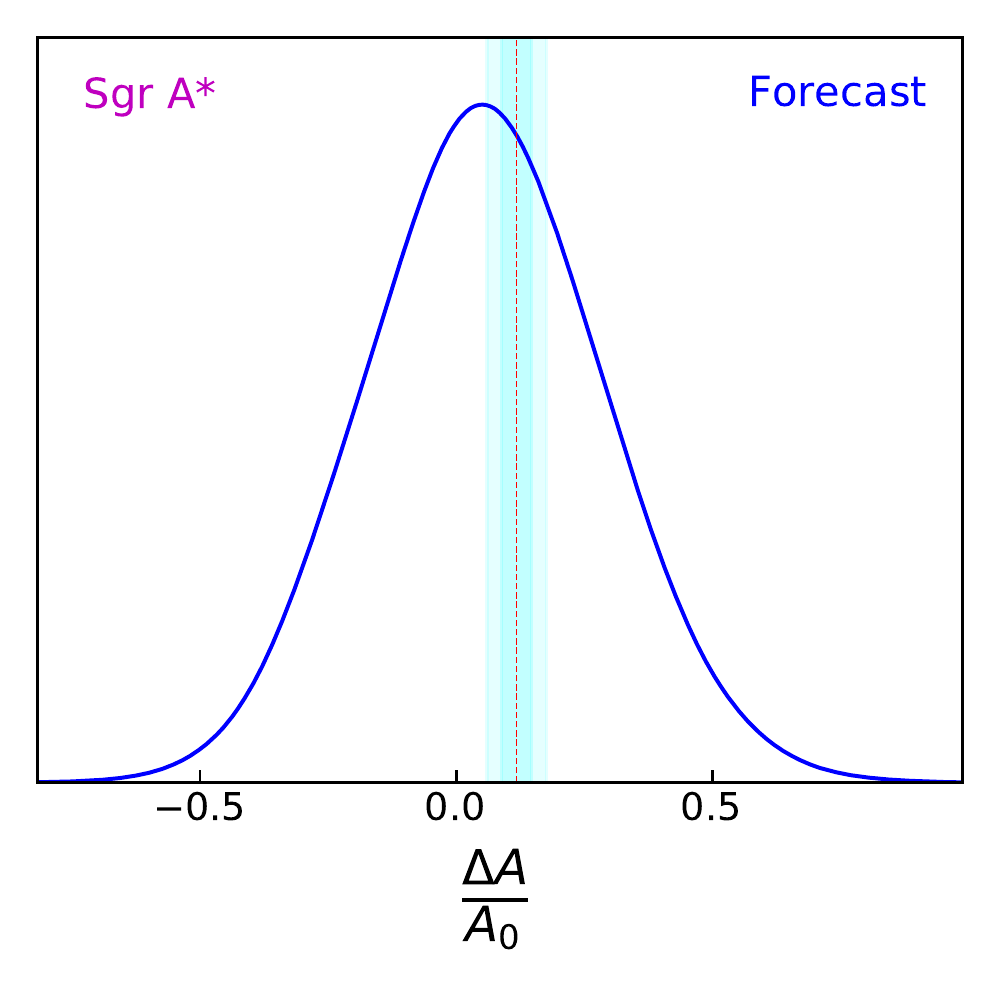}
	\caption{The 1-dimensional normalized posterior distributions of the forecasted area variation fraction $\Delta A/A_0$ of Sgr A* by improving the measurement precision of the angular ring diameter by 1 order of magnitude. The vertical red line and cyan bands are the mean and $1\,\sigma$ and $2\,\sigma$ confidence ranges of $\Delta A/A_0$, respectively.}\label{f4}
\end{figure}

\section{Methodology}
In general, an astrophysical black hole should be a Kerr one. To test the black hole area law, one should take the Kerr solution. However, the contribution to the event horizon radius from angular momentum decays rapidly to zero, since the EHT collaboration reconstructs the shadow images using the observations on our earth, which is far away from the observed supermassive black holes. Therefore, in the framework of GR, the shadow image should be consistent with the Schwarzschild black hole. In \cite{EventHorizonTelescope:2022xqj}, the EHT collaboration has tested the deviation from the Schwarzschild spacetime in light of their observations. It is safe to analyze the area variation of the supermassive black hole by using the Schwarzschild geometry. The horizon area of a Schwarzschild black hole with mass $M$ reads as  
\begin{equation}
A(M)=16\pi\left(\frac{GM}{c^2}\right)^2,    \label{1}
\end{equation}
where $G$ and $c$ are the Newtonian gravitational constant and speed of light, respectively. Subsequently, the area variation fraction of a black hole is defined as 
\begin{equation}
\epsilon\equiv\frac{\Delta A}{A_0}\equiv\frac{A_f-A_0}{A_0},    \label{2}
\end{equation}
where $\Delta A$ denotes the horizon area variation between the initial area $A_0$ and final area $A_f$. It is clear that the Hawking's area theorem requires $\Delta A/A_0\geq0$ during the evolution of a black hole. Furthermore, the angular ring diameter $d$ of a Schwarzschild black hole shadow is written as \cite{EventHorizonTelescope:2022xqj}
\begin{equation}
d = 6\sqrt{3}\,\theta_g\,\alpha_c\,(1+\delta),    \label{3}
\end{equation}  
where $\theta_g\equiv GM/c^2$ characterizes the gravity radius, $\alpha_c$ is the calibration factor and $\delta$ describes the deviation from the Schwarzschild black hole. By defining a correction factor $\alpha=\alpha_c\,(1+\delta)$ and combining Eq.(\ref{1}) with Eq.(\ref{3}), we obtain
\begin{equation}
A=\frac{4\pi}{27}\left(\frac{Dd}{\alpha}\right)^2,    \label{4}
\end{equation}
where $D$ is the distance from the source to our earth. One can easily see that the angular shadow radius determines the horizon area of a black hole when $D$ and $\alpha$ is given.

To confront the above model with observations, we implement the Bayesian statistics. To derive the likelihood function, we define $\chi^2$ as 
\begin{equation}
\chi^2=\left(\frac{\epsilon-\epsilon_{obs}}{\sigma_{obs}}\right)^2,    \label{5}
\end{equation}
where $\epsilon_{obs}$ and $\sigma_{obs}$ are the area variation and corresponding uncertainty derived from EHT data. We take a flat prior $\Delta A/A_0\in[-1,1]$ and carry out the Markov Chain Monte Carlo \cite{ForemanMackey:2012ig} analysis to obtain the posterior distributions of $\Delta A/A_0$. When implementing the numerical calculations, we use the SI units. To analyze the chains, we adopt the public code \texttt{Getdist} \cite{Lewis:2019xzd}.

\section{Data}
To perform the test of black hole area law, the angular ring diameters of supermassive black holes are needed. For M87*, we use the angular diameters $39.3\pm1.6$, $39.6\pm1.8$, $40.7\pm1.6$ and $41.0\pm1.4$ $\mu$as from the eht-imaging parameter survey for April 5, April 6, April 10 and April 11 in 2017 (see Fig.\ref{f1}), respectively \cite{EventHorizonTelescope:2019ths}. The distance from the earth to M87* is $16.8\pm0.8$ Mpc \cite{EventHorizonTelescope:2019dse}. We also take the correction factor $\alpha=11^{+0.5}_{-0.3}$ from the EHT analysis \cite{EventHorizonTelescope:2019dse}. For Sgr A*, similarly, we adopt the angular diameters $49\pm3.9$ $\mu$as and $50.0\pm2.5$ $\mu$as for April 6 and April 7 in 2017, respectively \cite{EventHorizonTelescope:2022wok}. The distance to Sgr A* is $8277\pm9\pm33$ pc measured by astrometric stellar dynamics \cite{Do:2019txf} and the correction factor $\alpha$ can be easily calculated by Eq.(\ref{3}) by using the data in Tab.I of Ref.\cite{EventHorizonTelescope:2022xnr}.

\section{Results}
By confronting the model with current EHT imaging observations, our results are presented in Figs.\ref{f2}-\ref{f4} and Tab.\ref{t1}. For M87* with mass $6.5\pm0.7\times10^9$ M$_\odot$ \cite{EventHorizonTelescope:2019dse}, we obtain the $1\,\sigma$ constraints 0.015$\pm$0.224, 0.060$\pm$0.230 and 0.015$\pm$0.215 on the horizon area variation fractions $\Delta A/A_0$ for three time intervals April 5 to 6, April 6 to 10 and April 10 to 11, respectively (see Fig.\ref{f2}). Across three mass orders, for Sgr A* with mass $4.297\pm0.013\times10^6$ M$_\odot$ \cite{Do:2019txf}, we have the $1\,\sigma$ constraint $\Delta A/A_0=0.118\pm0.232$ for the time duration April 6 to 7 in 2017 (see Fig.\ref{f3}). These constraints are consistent with Hawking's area theorem $\Delta A/A_0\geq0$ at the $1\,\sigma$ confidence level, since their error bars are large. Based on the fact that the measurement of gravity radius of Sgr A* from astrometric stellar orbit observations has a much better precision than that of M87* from the EHT-only data, we expect that whether the black hole area law is correct or not can be pinned down by future EHT observations of Sgr A*. As a consequently, we implement the forecast on the validity of the area law by roughly increasing the measurement precision of the angular ring diameter by 1 order of magnitude. This means that we shall adopt the angular ring diameters $49\pm0.39$ $\mu$as and $50.0\pm0.25$ $\mu$as for April 6 and April 7, respectively. We obtain $\Delta A/A_0=0.118\pm0.029$ (see Fig.\ref{f4}), which confirms the Hawking's black hole area theorem. 

Interestingly, it is easy to find that the mean values of $\Delta A/A_0$ for M87* and Sgr A* are all slightly larger than zero. This may indicates the correctness of the area law. Moreover, we are also interested in calculating the possibly realistic area variation rate $\Delta A/\Delta t$ and the Bekenstein-Hawking entropy variation fraction $\Delta S_{\mathrm{BH}}/S_{\mathrm{BH,0}}$. According to the best fitting values of $\Delta A/A_0$ in Tab.\ref{t1}, for M87*, we obtain $\Delta A/\Delta t=$ $0.81\times10^{21}$, $0.75\times10^{21}$ and $0.83\times10^{21}$, m$^2$\,s$^{-1}$,  $S_{\mathrm{BH}}/S_{\mathrm{BH,0}}=$ $1.77\times10^{-7}$, $1.63\times10^{-7}$ and $1.71\times10^{-7}$ s$^{-1}$ for three time intervals April 5 to 6, April 6 to 10 and April 10 to 11, respectively. For Sgr A*, we have $\Delta A/\Delta t=$ $0.82\times10^{21}$ m$^2$\,s$^{-1}$ and $S_{\mathrm{BH}}/S_{\mathrm{BH,0}}=$ $4.77\times10^{-7}$ s$^{-1}$ for the time duration April 6 to 7. One can find that the values of these two quantities for two supermassive black holes are of the same order, i.e., $\Delta A/\Delta t\sim\mathcal{O}(21)$ m$^2$\,s$^{-1}$ and $S_{\mathrm{BH}}/S_{\mathrm{BH,0}}\sim\mathcal{O}(-7)$ s$^{-1}$. This is because these two quantities are completely determined by the angular ring diameters, which for M87* and Sgr A* are of the same order.    

\section{Discussions and conclusions}
It has been a half century since Hawking's black hole area theorem was proposed. In light of the long-term monitoring observations of supermassive black holes M87* and Sgr A* from the EHT, we attempt to test the black hole area law.  

Using four and two black hole shadow images for M87* and Sgr A*, we obtain, respectively, the constraints on the horizon area variation fraction (see Tab.\ref{t1}). Our results are consistent with the prediction of Hawking's area theorem $\Delta A/A_0\geq0$ at the $1\,\sigma$ confidence level. However, due to large uncertainties of $\Delta A/A_0$ from current observations, we can not rule out the possibility that Hawking's area theorem may break down. Subsequently, by simply increasing the observational precision of the angular ring diameter by 1 order of magnitude, we find that future EHT observations of Sgr A* may confirm whether the black hole area law is valid or not. Furthermore, using the best fitting values of $\Delta A/A_0$, we also calculate the possibly realistic area variation rate $\Delta A/\Delta t$ and the Bekenstein-Hawking entropy variation fraction.

It is worth noting that the results are consistent but have a smaller uncertainty by a factor of $\sim$ 2, if not considering the effect of the correction factor $\alpha$ in Eq.(\ref{3}). This means that when roughly regarding the ring diameter as the shadow diameter, the constraint is stable and the corresponding uncertainty is suppressed. 

\section{Acknowledgments}
This work is supported by the National Science Foundation of China under Grants No.11988101 and No.11851301.

\end{document}